# A DM candidate indicated at Fermi-LAT and LZ ?

# Connection with LHC and LC prospects

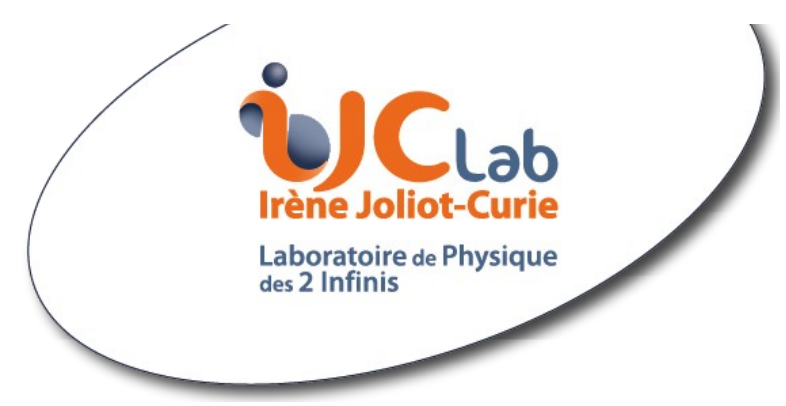

Alain Le Yaouanc[1], François Richard[2]

Université Paris-Saclay, CNRS/IN2P3, IJCLab, 91405 Orsay, France

September 12, 2026

Abstract

*Recently an analysis of* ***Fermi-LAT*** *data has claimed a significant excess of energetic photons centred around 20 GeV, in the* ***halo of our galaxy but absent in dwarf galaxies.*** *This excess can be interpreted as coming from dark mater (DM) objects with a* ***mass of few 100 GeV*** *which annihilate into hadronic jets containing photons mainly from neutral pions. This observation is independently backed up by a completely independent observation of the* ***LUX-ZEPLIN*** *(LZ) detector of a DM candidate in the same region of mass. To avoid contradiction with upper limits coming from* ***dwarf galaxies****, one is led to assume the presence of* ***a spin 2 resonance coupled*** *to these objects, which explains this difference as due to different speed distributions in the two regions***.** *This work recalls that LHC data indicate a* ***tower of Kaluza Klein graviton resonances****. This solution is at variance with a SUSY interpretation of DM and therefore does not necessary call for a 1.1 TeV Higgsino, as often proposed. If these KK resonances are coupled to e+e-, as indicated by LHC data, an* ***e+e- collider*** *could observe these DM objects.* ***LHC*** *could confirm our KK graviton interpretation and, eventually, detect the DM particles. This would allow to define de* ***energy needed at a future e+e- collider.***

1 Alain Le Yaouanc <alain.le-yaouanc@ijclab.in2p3.fr>
2 François Richard <francois.richard@ijclab.in2p3.fr>

# Introduction

DM matter can be detected either directly in Ar or Xe tanks or, indirectly, through photons coming from $\pi^0$ decays as suggested by the drawing below. For heavy DM particles, energetic neutrinos from leptonic decays originating from the sun can be detected by Ice Cube. DM can also be produced and observed at LHC.

A recent analysis of Fermi-LAT data [1] reports an excess of energetic photons, around 20 GeV, in the **halo of our galaxy.**

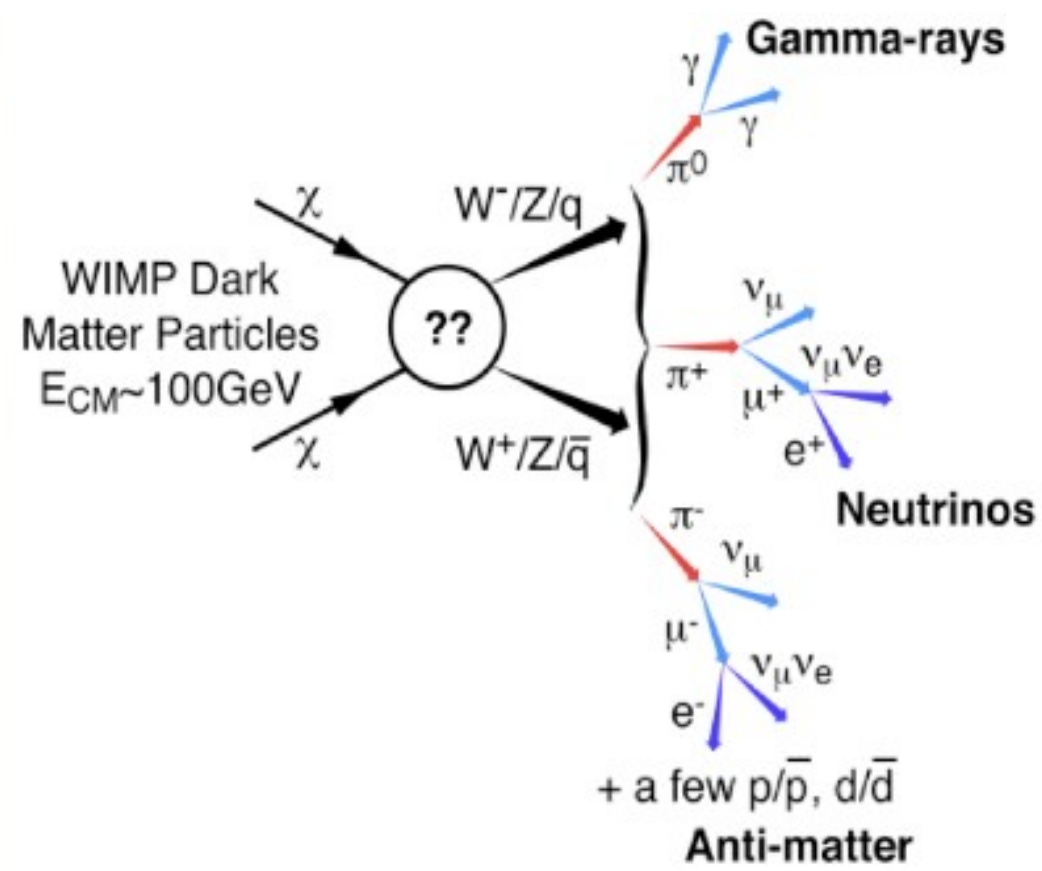

This can be interpreted as coming from dark mater (DM) objects with a mass of several 100 GeV, which annihilate into hadronic states with jets containing photons mainly from neutral pions.

Very recently the liquid Xe detector LUX-ZEPLIN (LZ) has observed a single DM candidate consistent with a similar mass [2]. This observation is significant, given that one observes a large energy deposit, well beyond what is expected for an elastic scattering, therefore almost free of background.

To avoid contradiction with upper limits coming from **dwarf galaxies**, [3] is led to assume the presence of **a ~TeV narrow resonance coupled near threshold for DM annihilation**. In dwarf galaxies the speed of DM particles is ten times lower than in our galaxy, meaning that for DM fermions annihilating into a J=2 resonance, the annihilation cross section is a hundred times lower than in dwarf galaxies.

# I. The LZ candidate

In this section we would like to recall an interpretation of the LZ candidate appearing in various recent references. This candidate is characterized by an energy deposit which makes it almost free of any known background, hence its high significance. To our knowledge there was no model predicting this effect. Several interpretations assume the presence of two Higgsinos states with a mass difference of order of the energy deposit seen in LZ. This interpretation does no seem to contradict the Fermi-LAT evidence since one expects that annihilation of such DM particles will take place giving WW/ZZ final states producing energic photons originating from hadronic jets. SUSY models predict that this particle is a Higgsino which could be highly degenerate in mass with the lightest chargino and therefore difficult to observe at LHC.

Note that several authors underline that DM annihilation in the sun should produce energetic neutrinos which should be detected by Ice cube [13]. Note however that this requires very energetic neutrinos, hence very massive DM which may not be the case. [14] claims a mechanism which could weaken this contribution.

# II. Interpreting the Fermi-LAT result

Figure 1 shows the photon spectrum in excesss to the predicted background observed in an analysis of Fermi-LAT data. It can be reproduced assuming that DM particles annihilate primarily into ZZ/WW. It could also annihilate into two photons which should a peak at the tail of this distribution. The energy resolution of Fermi-LAT is ~10%. ThIs is consistent with a slight excess observed around 200 GeV in figure 1. This effect, still marginal, would correspond to a DM particle with mass ~200 GeV.

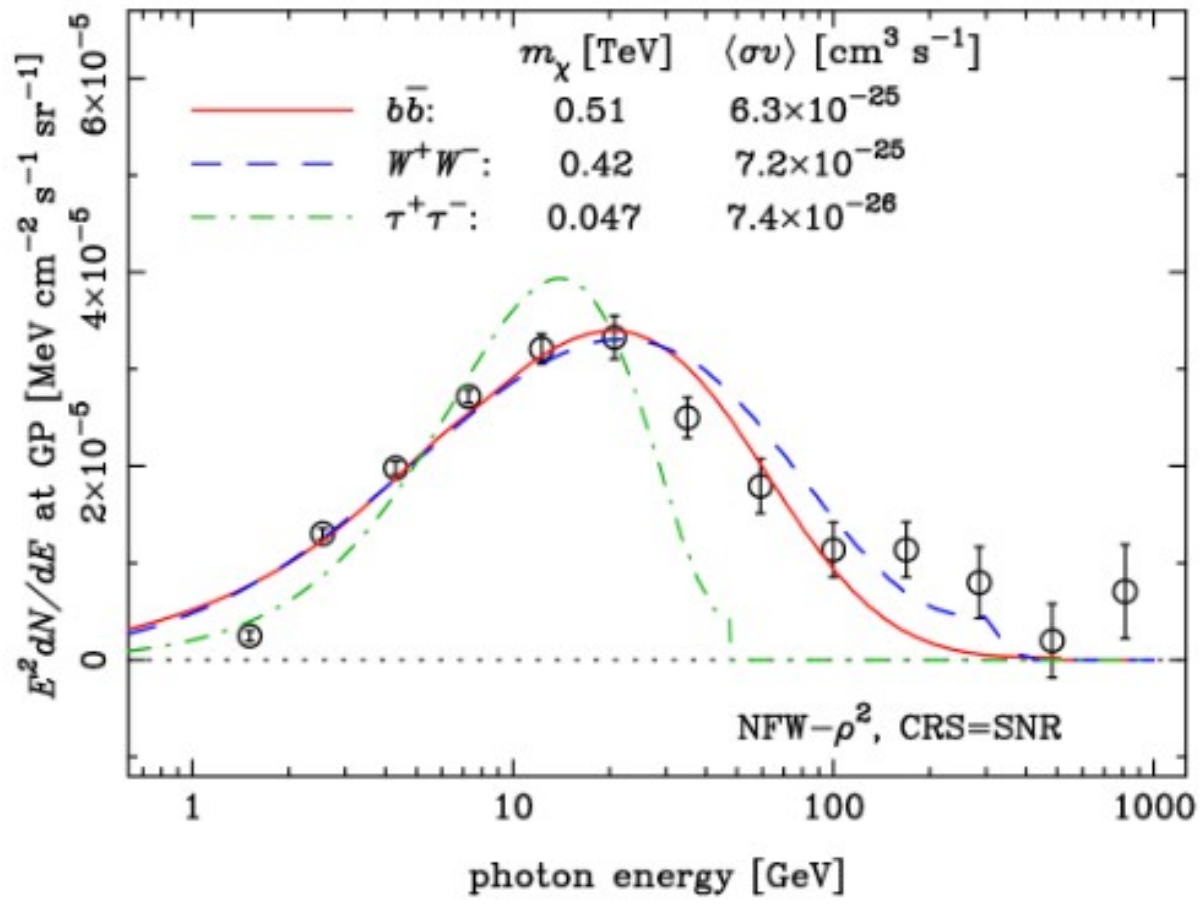

*Figure 1:Energy distribution of the excess of photons observed in [1].*

In reference [3] one assumes that these annihilation take place through a narrow resonance which comforts the idea that there could be a photon peak near the maximum if this resonance decays at the % level into two photons.

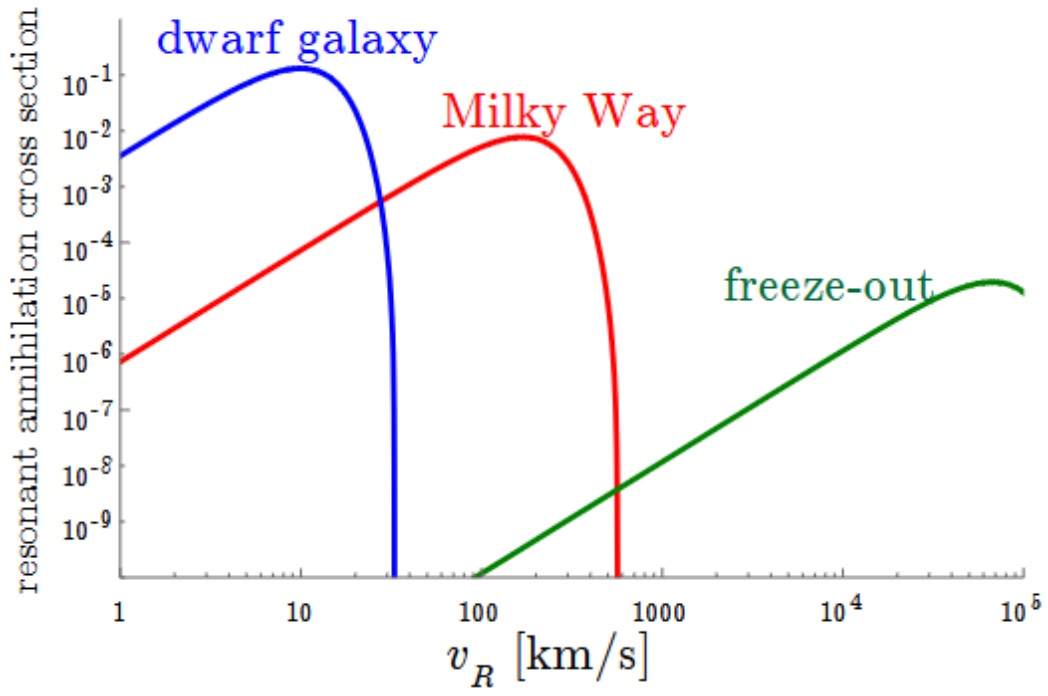

*Figure 2:Resonnant annihilation cross section versus the speed of DM particles assuming annihilation through a scalar resonance [3].*

Figure 2 shows the expected resonant annihilation cross section assuming a J=0 resonance. It does not explain the absence of signal from the dwark galaxies. If, instead, one assumes that one has a J=2 resonance, one should have a **p-wave dependence** on this cross section which would simply explain the absence of signal for dwarf galaxies.

This feature considerably reinforces the importance of an indication for a **tower of spin 2 KK resonances** observed by LHC which is described in [4,5]. One of these resonances should play a capital role in providing an adequate resonant annihilation of DM particles in our Universe. This annihilation needs not take place right at the top of this resonance, which would be reasonably wide (see figure 3) meaning that this interpretation does not suffer from fine tuning.

An other important conclusion that one can draw is that this mechanism of annihilation through spin 2 resonances differs radically from the MSSM case for Higgsinos which proceeds through scalars. For this reason one cannot support the **usual assumption that the DM is a Higgsino with a 1.1 TeV mass.**

One should therefore consider **alternate DM interpretations** as proposed in extra-dimension models which would match with the presence of KK graviton resonances (see for instance [6]).

In the next section we will recall the evidence for KK graviton resonances observed at LHC.

# III. Interpreting LHC searches

LHC searches are relevant in several ways. LHC has searched for charginos and neutralinos which are plausible candidates for DM. We will briefly recall present results and show that they are reaching a mass region relevant for our case.

We will first recall our results showing some indication for a tower of KK gravitons which could also be a BSM source for DM particles since such resonances can abundantly couple to pairs of charginos.

## III.1 Evidence for KK gravitons

In [4,5], we claim that LHC data indicate the presence of Kaluza Klein graviton resonances, few of them which could do the job for DM annihilation. This work has collected the various available indications for resonances at LHC and concluded to the presence of a sequence of resonances which starts at 380 GeV followed by 700 GeV and 1 TeV. This sequence is quantitatively understood within the **Randall Sundrum model.** Next resonances are predicted at 1300 and 1620 GeV. The third resonance is indicated by a search for Higgs pairs at 1 TeV as reported in [4,5].
The various observed states are displayed in figure 4 while the following table display the various observations.

**4 leptons $\gamma\gamma$** **hh** **ZZ semi-lept VBF** **ZZ+WW had**

| m $G_{KK}$ GeV | 380 | **700±20** | 1000 | 1322 | 1640 | 1945 | 2260 | 2590 | 2905 |
|---|---|---|---|---|---|---|---|---|---|
| $\Gamma$ $G_{KK}$ GeV | 3 | **20±7** | 60 | 130 | 260 | 440 | 680 | 1042 | 1440 |
| VBF→$G_{KK}$ fb | 360 | **100±50** | 38 | 15 | 7 | 3.4 | 1.7 | 0.9 | 0.4 |
| e+e-→ $G_{KK}$ pb | 450 | **130** | 60 | 50 | | | | | |

In the Randall Sundrum model, $G_{iKK}$ resonances have **quantified masses** [7] which, in our case, go like $m_i \sim 100 x^G_i$ in GeV, where: $x^G_i$=3.83,7.02,10.17,13.32,16.5... The total width itself goes like $\Gamma_i \sim 0.06(x^G_i)^3$ in GeV.

The table above summarises these findings together with some estimates of the LHC cross sections. In contrast to the scalar resonances predicted by extended Higgs models, the first KK graviton resonances are predicted narrow and likely to be observed in the two photon mode. The following two figures show the evidence for the second KK graviton resonance in two photons, e+e- and ZZ→4 leptons modes. While the two first modes are obviously narrow, the ZZ mode is distorted by an interference with the SM background. When this effect is properly taken into account, the resulting width appears compatible with the two others.

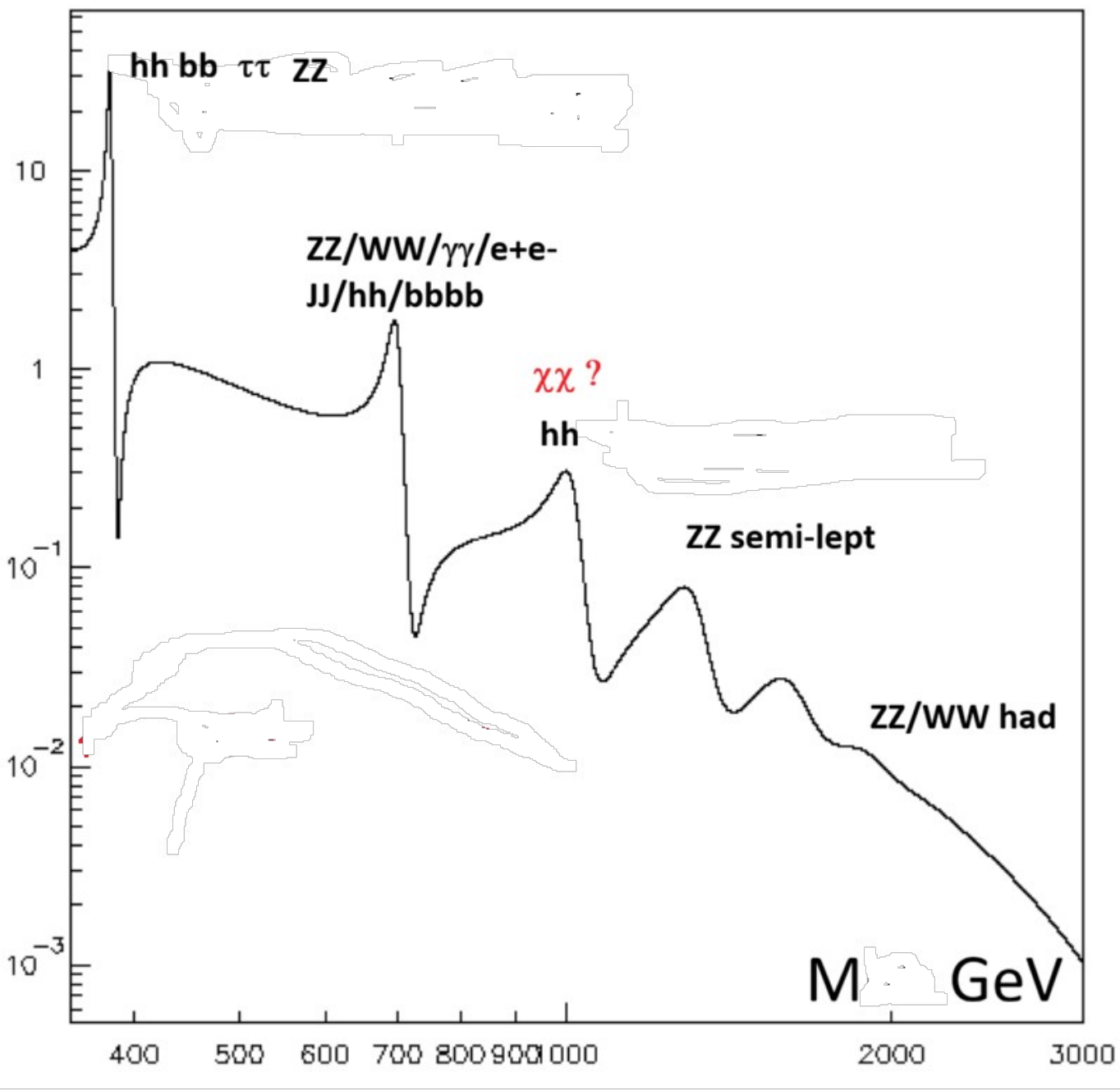

*Figure 3:From [4] variation of the KK-graviton cross section versus the masses of the KK rresonances. Indicated are the channels showing excesses.*

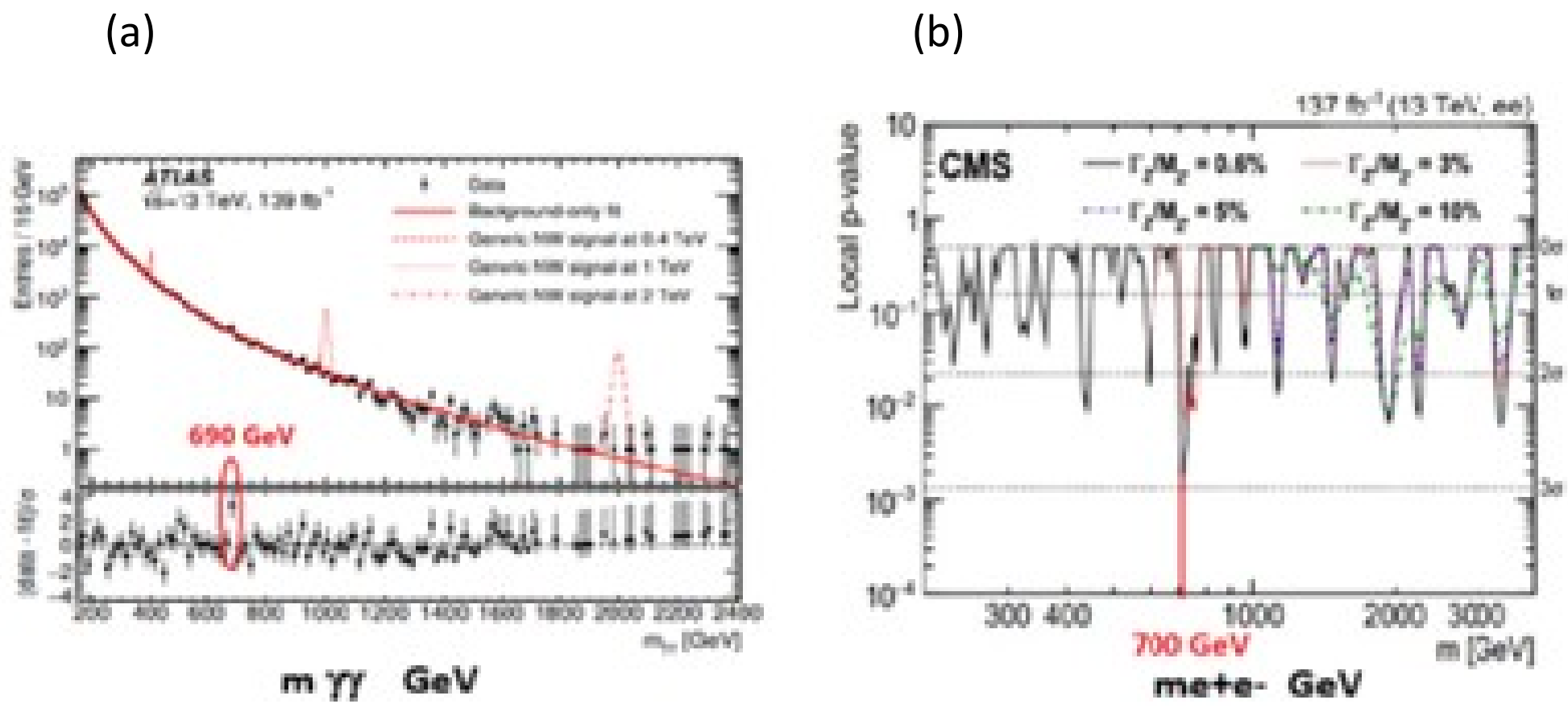

*Figure 4: From [4] two photon (a) and e+e- (b) mass distributions observed by ATLAS and CMS indicating a narrow peak at a mass close to 700 GeV.*

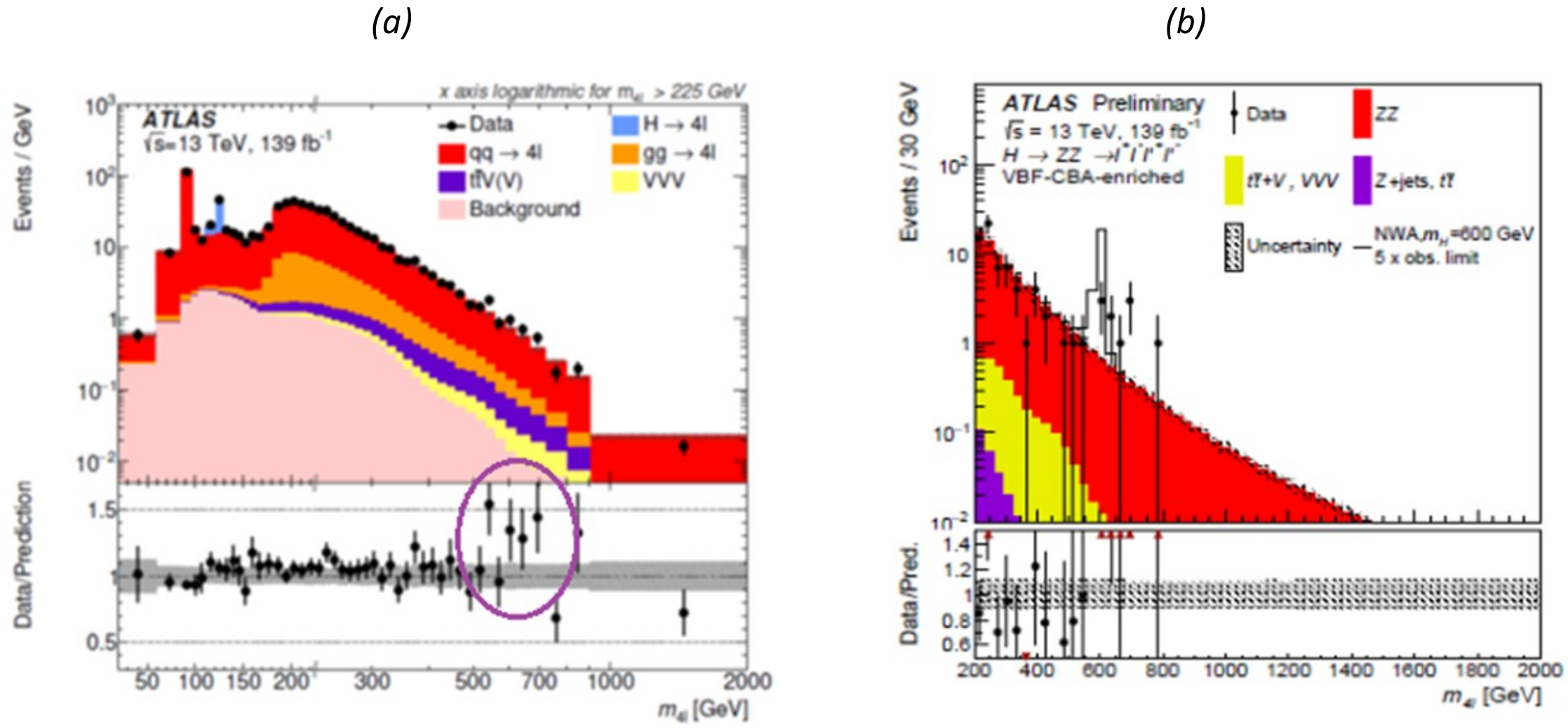

*Figure 5: From [4 and [5], the ZZ mass distribution observed in 4 leptons by ATLAS indicating a structure around 700 GeV. Plot (a) is cut based analysis for ggF and (b) for VBF.*

An additional proof of our KK graviton interpretation comes from the following observation. This indication is only significant when applying genuine selections while it disappears if one applies drastic cuts which assume an isotropic angular distribution as would be the case for a scalar resonance. A J=2 resonance produced in the VBF mode, which seems the case for this resonance, is indeed forward peaked as it is the case for the Drell Yan background.

## III.1 Searches for charginos at LHC and their limitations

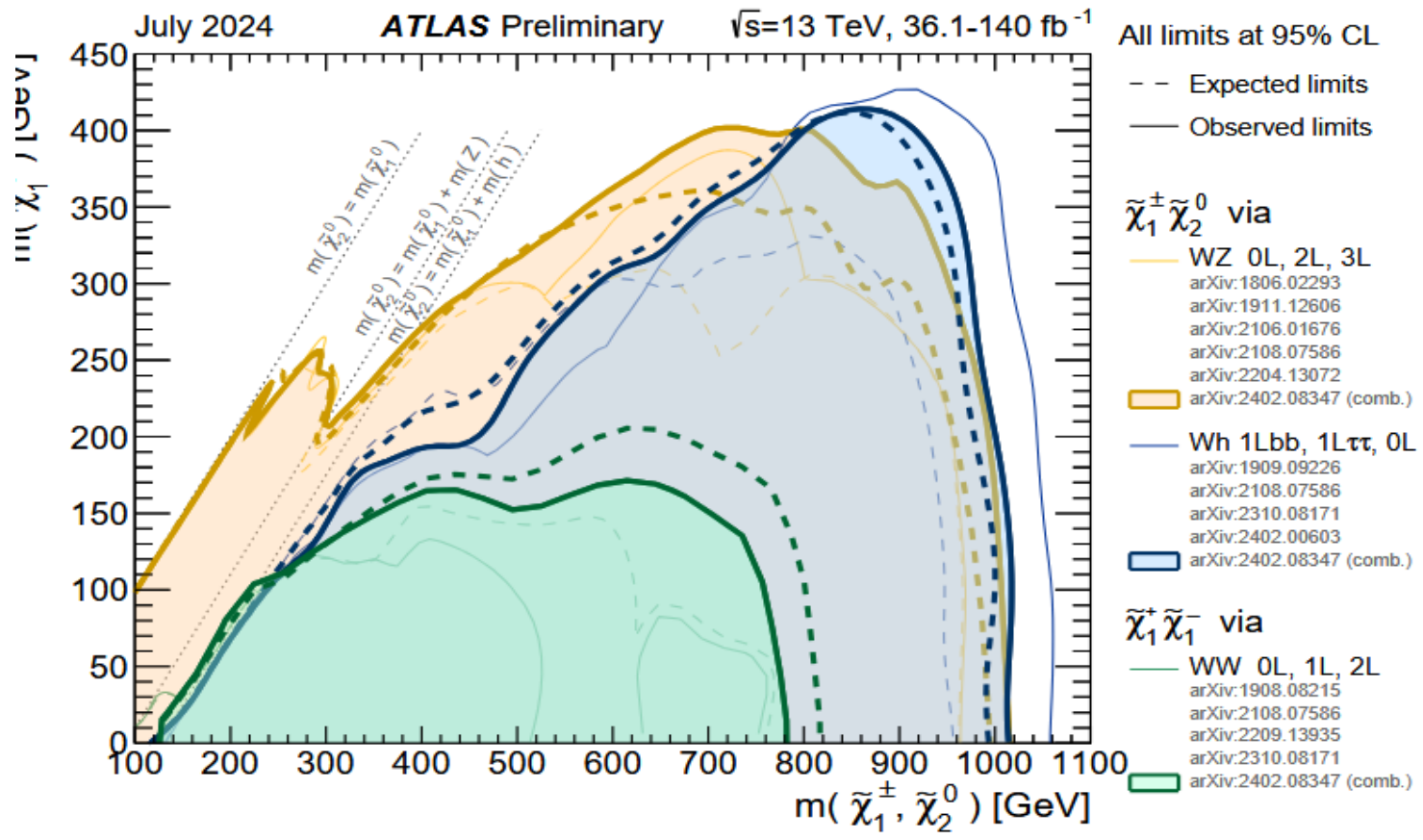

*Figure 6 from [8] shows the 95% CL exclusions limits for chargino-neutralino and chargino pairs searches assuming a pure wino solution.*

As an example [8] figure 6 shows that an exclusion reaches 400 GeV for LHC searches for the wino case. In case of a Higgsino with mass degeneracy between the chargino and the LSP, exclusion only reaches 200 GeV [9].

Some excesses are reported in LHC data, still marginally significant. Reference [10] interprets these excesses as coming from a chargino with mass below 250 GeV.

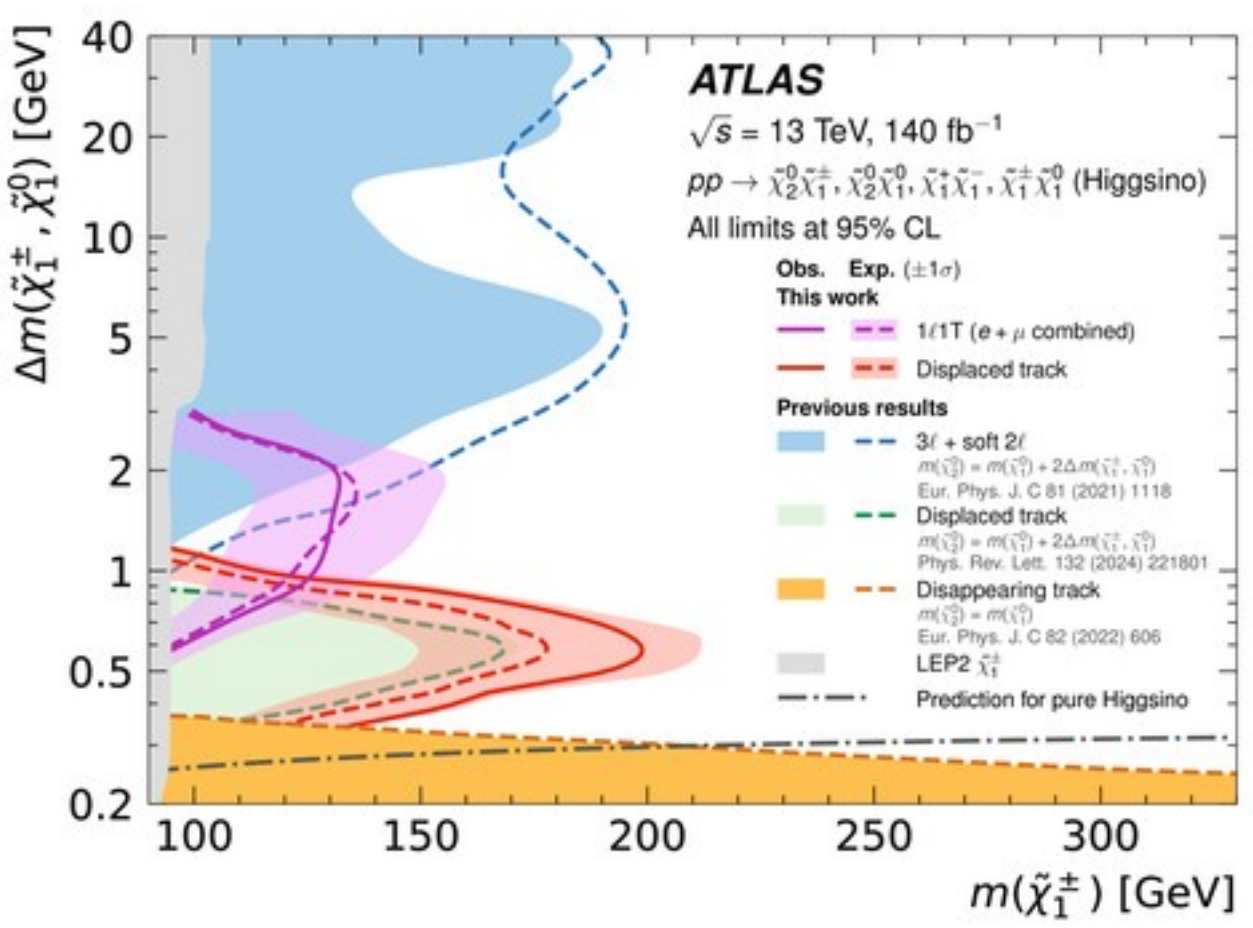

Figure 7: Status of ATLAS searches for chargino-neutralino quasi-degenerate in mass [10].

For the highly degenerate case one selects ‘disappearing tracks’ [11]. It reaches mass limits up to 700 GeV.

# IV. Predictions for an e+e- LC

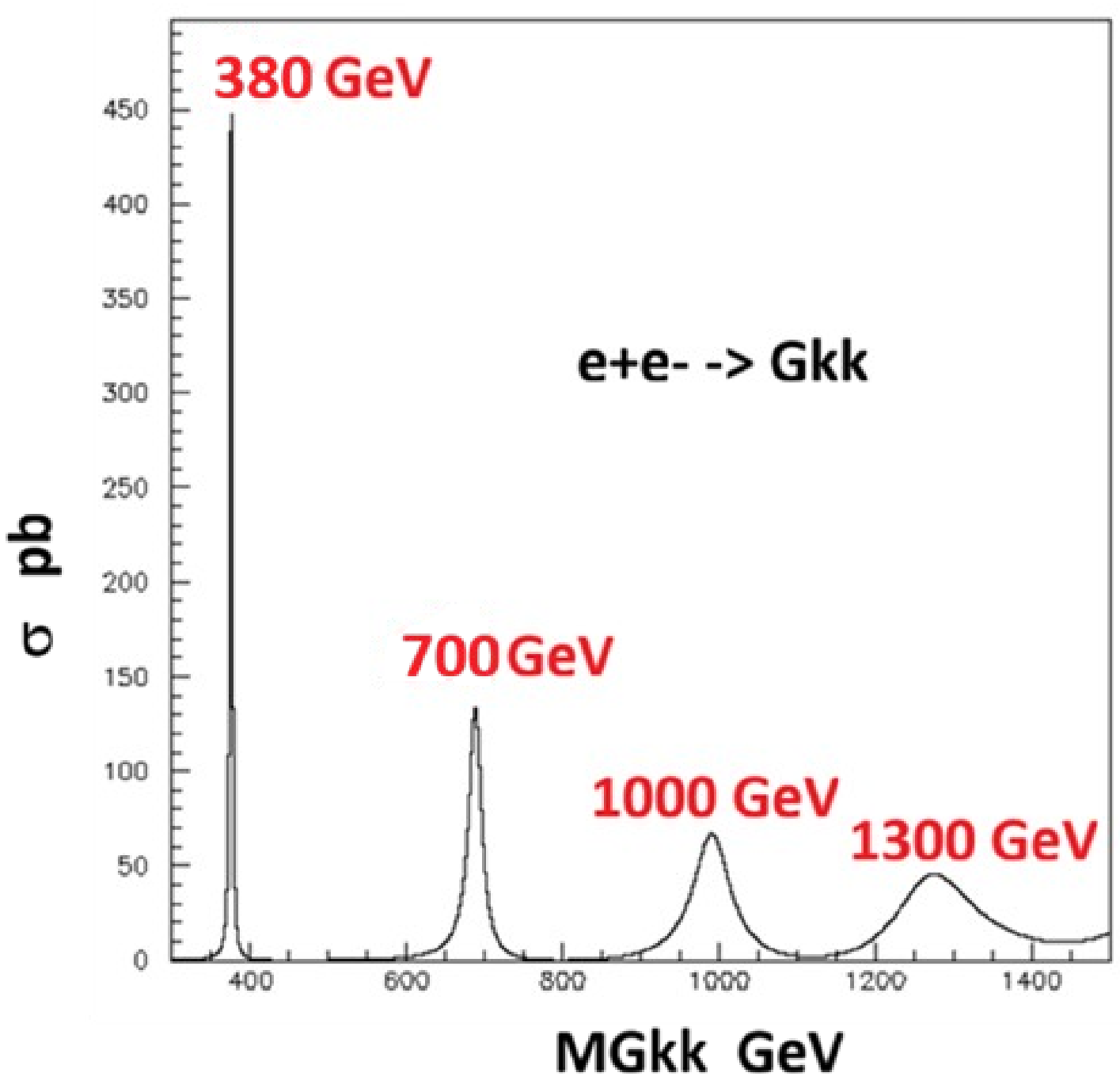

*Figure 8: e+e- cross sections for Kaluza Klein graviton resonances [4,5].*

Figure 8 shows the predicted cross sections for the various KK graviton resonances in an e+e- collider. They are huge, recalling, for instance that the ILC TDR assumes for 1 TeV an integrated luminosity ~8000 fb-1, hence 0.5 Giga-events.

To understand cosmology one needs to measure the coupling of DM to KK gravitons. One can conceive two scenarios.

1/ The standard method used at the Z pole one measures accurately the invisible width. It turns out however that on top of the **threshold resonance $G_{KK}$i**, one sits very near threshold for DM pairs, which leads to a negligible invisible BR with respect to the SM contribution from ZZ into neutrinos which is of order 1%. One therefore needs to operate on the **i+1** recurrence.

2/ Assuming that DM has charged partners, one can produce an abundant set of such particles by operating at the **i+1** recurrence, which will allow to determine the properties of these particles for a SUSY case.

In the case of SUSY, BR of $G_{KK}$ into charginos is expected to be substantial as discussed in the next section.

## III.1 Predictions for charginos

Let's assume a SUSY interpretation.

| $MG_{KK}$i GeV | 700 | 1000 | 1300 |
|---|---|---|---|
| $M\chi+$ GeV | 190 | 350 | 500 |
| $e+e \to G_{KK}$ fb | 130000 | 70000 | 50000 |
| $BR(KK \to \chi+\chi-)$ % | 34 | 31 | 21 |
| $e+e- \to G_{KK}i \to \chi+\chi-$ fb | 45000 | 22000 | 10500 |
| $e+e- \to \gamma \to \chi+\chi-$ fb | 171 | 70 | 34 |

In principle [3] allows to compute the decay partial width into DM for the active resonance (which one, we cannot tell a priori), correcting the formulae to take into account a spin 2 solution. There is however a caveat: [3] assumes that 'nature' chooses DM masses so that annihilation takes place **at the top of one of these resonances.** This may not be the case and, judging from figure 8, one sees that this introduces very large uncertainties, specially for the first resonances which do not overlap. For this reason we have refrained from predicting the BR of these resonances into a pair of DM.

An e+e- collider produces charginos with, at minimum, a point like cross sections:

$$\sigma_{fb}=100\beta(1+\beta^3/3)/s_{TeV^2} \quad \beta=\sqrt{(1-4m^2/s)}$$

$$\chi+\chi-/W+W-=3/2(1-4x^2)^{3/2}(1+8x^2/3) \qquad \text{where } x=m\chi+/mG_{KK}$$

# Conclusion

LZ has published a DM candidate which seems almost background free. This result was based on 200 days pf data taking and the plan is to reach 1000 days. If the DM interpretation is true, this experiment has good chances to fulfill the discovery criterion. These results seem to coincide with an evidence claimed after analyzing publicly available results from Fermi-LAT.

Fermi-LAT provides an essential information concerning the annihilation mechanism for DM. The absence of a contribution of dwarf galaxies suggests that annihilations take place in an **p-wave** which is contrary to the usual MSSM assumption of Higgsinos coupled to scalar particles. One therefore cannot draw the usual conclusion that we are dealing with 1.1 TeV Higgsino since this assumes an annihilation through a scalar resonance in an **s-wave.** One should also remain agnostic for what concerns a standard SUSY interpretation of DM particles and consider **alternative sources of DM as proposed in the RS model**.

The photon spectrum shows a weak indication for a small bump at 200 GeV which suggests a decay of a resonance $G_{KK}$(380) into two photons and a mass for DM of 190 GeV. More data and an improved energy resolution would be welcome.

HL-LHC should be able to give very important inputs, either discovering the DM particle or, at least, setting some lower bound on its mass and confirming the KK graviton interpretation. At the moment LHC provides a small indication for a 200 GeV chargino type particle. HL-LHC should also provide an other essential information: the size of the coupling of the KK graviton resonances to e+e-, necessary to guarantee the opportunity of e+e- KK graviton factories.

In a bottom up approach, it seems that we are gradually building a model comprising extra dimensions and, perhaps, SUSY. If confirmed, this provides a very rich description of our world, incorporating the ingredients of a TOE, the so-called super strings where SUSY and extra dimensions appear in conjunction. The next years will tell what remains from this very rich landscape and how particle physics can further develop with a new collider.

**Acknowledgements:**
*Thanks to **Akira Miyazaki** for drawing our attention on the paper of Hitoshi Murayama about DM searches in Fermi-LAT. We owe him many useful inputs and lively discussions.*
*Thanks to **François Couchot** for stimulating discussions.*
*Many thanks to **Hitoshi Murayama** for useful exchanges on the Fermi-LAT results.*
*We are grateful to **Jenny List** for encouraging the presentation of this work at LCWS2026*

# APPENDIX

## Beyond a Randall Sundrum interpretation of KK gravitons

In the most naive interpretation of the Randall Sundrum model, precision measurements seem to exclude KK particles below 30 TeV. Extensions of the model with higher symmetries (presence of Z') allow masses down to 2 TeV.

Other extensions allow wide hierarchies between different types of KK particles. One radical example of this is shown below :

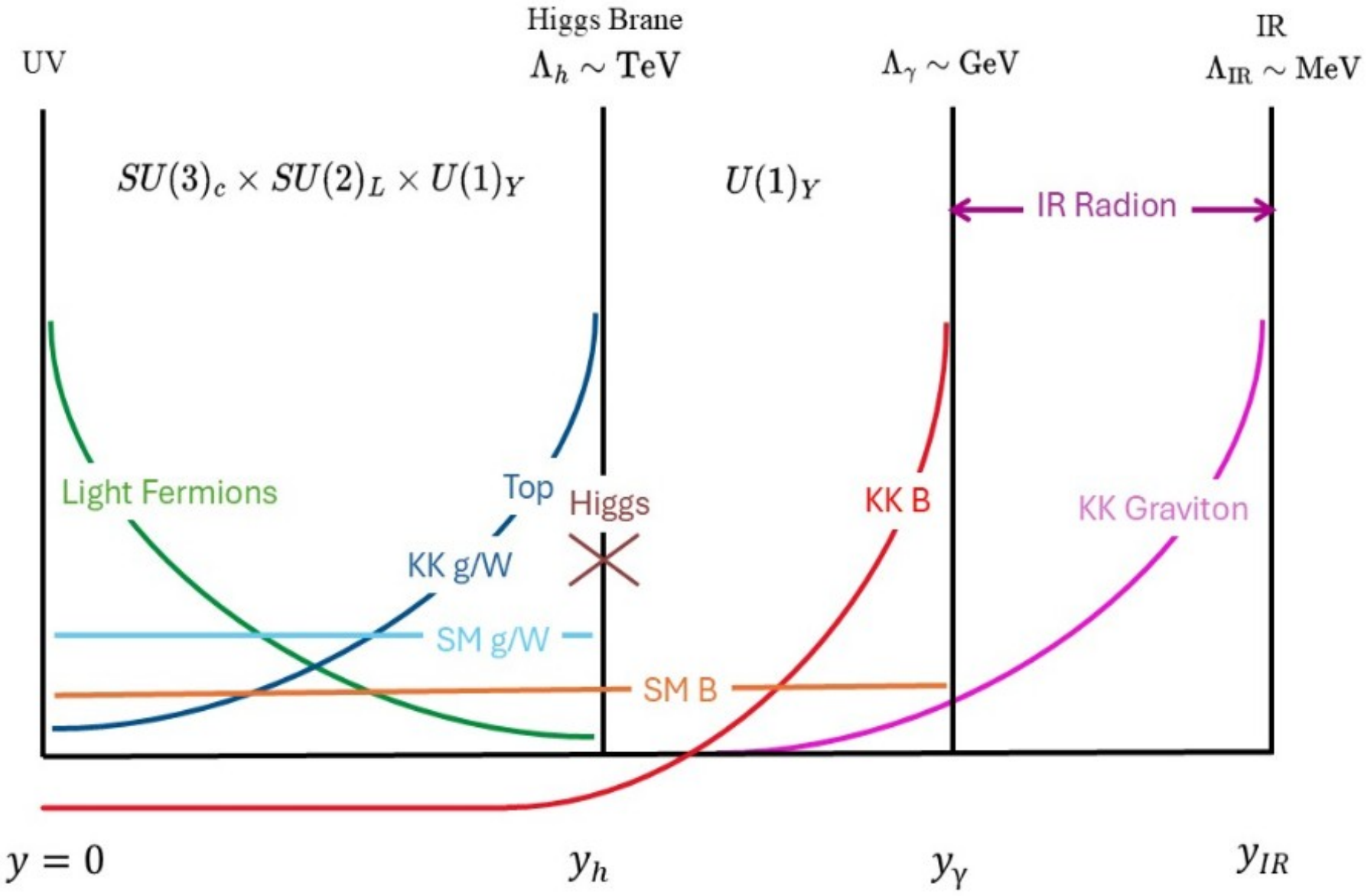

*Figure 9: [12] shows an extended version of the Randall Sundrum model with additional branes which allows non universality behaviour of the various KK recurrences of SM particles.*

where the KK gravitons can become much lighter than other KK particles [12].
Experimentally speaking, the standard RS picture which assumes that KK gravitons **couple directly** to gluons and photons implies an exclusion up to ~2 TeV.
These direct couplings do not exist for the SM Higgs where they take place uniquely through tt and WW loops.
To interpret this, the **dual composite model** assumes that the preons composing the Higgs particle carry no charge and no color.

Why could't this be true for KK gravitons ? In which case the presence of KK graviton resonances would still be allowed below a TeV by LHC searches.